\documentstyle[12pt]{article}
\newcommand{\beq}{\begin{equation}}
\newcommand{\eeq}{\end{equation}}
\newcommand{\nn}{\partial\hspace{-.55 em}/}
\newcommand{\dd}{D\hspace{-.75 em}/}
\newcommand{\h}{\theta}
\newcommand{\r}{\varphi}

\newcommand{\e}{\bar{\eta}}
\newcommand{\p}{\bar{\psi}}
\newcommand{\s}{\varepsilon}
\newcommand{\gv}{\rm GeV}
\def\tt{$\theta$-term}
\def\rt{rotation}
\def\trn{transformation}

\def\zm{zero mode}
\def\op{operator}
\def\f{field}
\def\lr{$SU(2)_L \times SU(2)_R$}
\def\cn{condition}
\begin{document}
\titlepage

\title{CAN ELECTRO-WEAK $\h$-TERM BE OBSERVABLE ?}
\date{}
\author{A.A.Anselm and A.A.Johansen
\thanks{On leave of absence from Petersburg Nuclear Physics Institute,
Gatchina, 188350, St.Petersburg, Russia}\\
Petersburg Nuclear Physics Institute\\
Gatchina, 188350 St.Petersburg, Russia}
\maketitle

\begin{abstract}
We rederive and discuss the result of the previous paper that in the standard
model $\theta$-term related to $W$-boson field can not be induced by weak
instantons.  This follows from the existence of the fermion zero mode in the
instanton field even when Yukawa couplings are switched on and there are
no massless particles.  We consider the new index theorem connecting the
topological charge of the weak gauge field with the number of fermion zero
modes of a certain differential operator which depends not only on gauge
but also on Higgs fields.  The possible generalizations of the standard
model are discussed which lead to nonvanishing weak $\theta$-term. In
$SU(2)_L \times SU(2)_R$ model the $\theta$ dependence of the vacuum energy
is computed.
\end{abstract}

\newpage
\section{Introduction}

In recent years much attention has been paid to the nontrivial topological
fluctuations of the weak gauge field (W-boson \f ).
One of the main reasons for that is that the baryon number is violated
through these fluctuations due to the axial anomaly in the
divergence of the baryon current \cite{1}.
The interest to this problem has been particularly enhanced
when it was argued that the exponential suppression of baryon
non-conservation may be absent at high energies of order of $30-50$
TeV \cite{ring}.

An interesting question which can be raised in relation to this
observation is the existence of the "weak $\theta$-term"
which would violate CP-invariance in the processes with baryon
non-conservation.
By the "weak $\theta$-term" we mean the following
contribution to the action:
\beq
\Delta S = \theta \frac{g^2_W}{32\pi^2}\; \int d^4 x W^a_{\mu\nu}
\widetilde{W}^a_{\mu\nu}.
\eeq
Here $W^a_{\mu\nu}$ is the field strength constructed from the
$W$-boson field,
$W^a_{\mu\nu} =
\partial_{\mu} W^a_{\nu}-
\partial_{\nu} W^a_{\mu} + g_W \epsilon_{abc} W^b_{\mu} W^c_{\nu}$,
$\widetilde{W}^a_{\mu\nu} = 1/2 \; \epsilon_{\mu\nu\lambda\sigma}
W^a_{\lambda\sigma}$, $g_W$ is the weak gauge coupling
constant and $\theta$ is a constant parameter.

Similar to the "strong $\theta$-term", constructed
of the gluon field, the weak $\theta$-term
could appear in the action as a result of a diagonalization
of the fermion mass matrix.
As is well-known this diagonalization requires chiral rotations
of fermions.
The chiral rotations correspond to anomalous symmetry
transformations and, therefore,
result in the appearence of the \tt \ (1) in the action.
Note that in the case of the weak \tt \ both quarks and leptons
contribute to (1) contrary to the case of strong \tt \ where
only quarks are involved.

Another reason why one should worry about the weak \tt \
is that it can be added to the action "by hands",
being renormalizable and conserving all the symmetries of the
theory.

The existence of the weak \tt \ would lead to CP-violation in the
processes with baryon non-conservation.
That could have been important for cosmology.

The investigation which we have carried out in the previous paper
has shown, however, that the weak \tt \ is
not observable in the standard model of electro-weak interactions
\cite{aj}.
It turns out that since the masses of fermions are generated by
spontaneous symmetry breaking the situation is similar to that of
the strong \tt \ in the presence of massless quarks,
where the \tt \ is unobservable.

In the latter case vanishing of the strong \tt \ in the action is
due to the existence of the \zm \
of the massless quark
in the instanton gluon field.
The same arguments have been used in ref.\cite{aj} to show the
vanishing of the weak \tt .
The existence of the \zm \ for massive fermions in weak instanton field
has been first discovered in ref. \cite{rub}.

It is convenient to describe the problem of weak \tt \ using
the generalization of the $\gamma_5$ invariance,
the $\Gamma_5$ symmetry, which has been introduced in ref.\cite{aj}.
The Lagrangian of the weak interactions including the
Yukawa interactions is $\Gamma_5$ invariant
since $\Gamma_5$ symmetry trivially reduces in this case
to the simple (nonchiral) phase transformations.
However the notion of $\Gamma_5$ symmetry proves to be useful in
order to formulate a new index theorem for an \op \ generalizing
the Dirac \op \ $\dd = \gamma_{\mu} D_{\mu}$ for the presence of the
Yukawa couplings.
In this context $\Gamma_5$ symmetry acts rather as a chiral than as a
phase transformation.
In turn, the index theorem allows to establish the existence
of the massive fermion \zm \ in the instanton field mentioned above.

In this paper we first consider the problem of the weak \tt \ from
somewhat different point of view than it has been done in ref. \cite{aj}.
We rederive the existence of the fermion \zm \ and the index theorem
proven in ref. \cite{aj}.

Then we pass to the main goal of this paper: to find out the
\cn s under which the weak \tt \ can exist and be observable.
These \cn s lie beyond the standard model and we consider two
possibilities.
One is an additional Yukawa type interaction explicitly violating baryon
number.
Another is the left-right symmetric generalization
of the standard electro-weak group.

In the first case CP-violation, related to the weak \tt ,
is suppressed by a small Yukawa coupling constant of the
interaction violating baryon number, in addition
to the exponential suppression
$exp(-8\pi^2 /g^2_W)$ charachteristic for the instanton induced amplitudes.
In the second case there is a double exponential suppression:
$exp(-8\pi^2 (1/g^2_L + 1/g^2_R))$, where $g_L = g_W$ is the usual
weak coupling constant while $g_R$ is the gauge coupling of the
additional $SU(2)_R$ group.
In the latter case one might expect the vanishing of the
exponential suppression at energies higher than
$8\pi^2 M(W_R)/g^2_R$ ($M(W_R)$ is the mass of
the $SU(2)_R$ gauge bosons)
by analogy to the standard model \cite{ring}.

The qualitative explanation of the appearence of the \tt \ in
the \lr \ model is the following.
In the standard model, with weakly interacting left-handed
fermions, the \tt \ (1) can not be observable since the rotations
of the left-handed fermions would shift $\theta$ by an arbitrary
constant due to the axial anomaly.
The rotations of the left-handed fermions can be compensated
here by rotations of right-handed fermions to make the Yukawa couplings
invariant to this transformation.
Of course, the real cause why \tt \ (1) is unobservable in this case
is that for non-trivial topological fluctuations of the $W$-boson \f ,
when the integral (1) does not vanish,
the contribution to the partition function from these fluctuatioins
vanishes due to the fermion \zm . (See below).

In the case of the \lr \
model the rotation of right-handed fermions,
which is necessary to provide the invariance of the Yukawa couplings,
induces the \tt \ similar to that of eq.(1) but with $W_{\mu\nu}^a$
changed to the gauge \f s of the $SU(2)_R$ group and of the opposite
sign. If the topological charges in $L$- and $R$-sectors are
the same two anomalies cancel each other and the initial \tt \ can not
be rotated out.
This implies that the fermion zero mode disappears,
the contribution of these gauge \f \ fluctuations to the
partition function does not vanish and the \tt \ becomes observable.

\section{Weak $\h$-Term in Standard Model}

We consider the simplified version of the standard model,
with the weak hypercharge and colour interactions omitted since they
are irrelevant to the discussion below.
For simplicity we restrict ourselves for the moment
by consideration of two weak left-handed chiral fermion doublets,
for instance:
\beq
q_L\; =\; \left( \begin{array}{c} u\\ d\end{array}\right)_L, \qquad
q'_L\;=\;\left( \begin{array}{c} u'\\ d' \end{array} \right)_L.
\eeq
Here $q_L$ and $q_{L'}$ can be either quark doublets with different
colour indices, or they may mean quark and lepton doublets.
Correspondingly we introduce the singlet right-handed chiral
fermions $u_R, d_R, u'_R, d'_R$.
The generalization for the case of all twelve existing doublets is
straightforward.

When the weak doublets are considered in pairs it is possible to pass to the
pure vector interaction of fermions with $W$ bosons. To do this we introduce
instead of fermion fields $q'$ the charge-conjugated fields:
\beq
\tilde{q}'_R =\,\s C\bar{q}'_L = \left( \begin{array}{c} C\bar{d}'_L\\
-C\bar{u}'_L \end{array}\right), \qquad \tilde{q}'_L=\s
Cq'_R=\left(\begin{array}{c} C\bar{d }'_R \\ -C\bar{u}'_R \end{array}\right).
\eeq
Here $\s=i\sigma_2$ acts on the isotopic indices, $C$ is the
charge-conjugation matrix. If now one composes two Dirac spinors
\beq
\begin{array}{lll} \psi=\psi_L+\psi_R,        &  \psi_L=q_L,   &
\psi_R=C\s\bar{q}'_L \\  \eta=\eta_R+\eta_L, & \eta_R=q_R, &
\eta_L=C\s\bar{q}'_R, \end{array}
\eeq
then it is obvious that both components
of $\psi,\, \psi_L$ and $\psi_R$, are weak doublets while $\eta_R$ and
$\eta_L$ are singlets. Therefore only the $\psi$ field has a vector gauge
interaction:
\beq
{\it L}_W=i\p \dd\psi + i\e\nn\eta, \qquad  \dd=\gamma_\mu D_\mu=\gamma_\mu
(\partial_\mu-i\hat{W_\mu}).
\eeq
Clearly the mixing of the quark and antiquark fields in eq.(4) is very
unnatural with respect to colour and electric charge (weak hypercharge).
However
as it has been already mentioned
colour interaction and interaction with the weak hypercharge are
irrelevant to the problem considered.

In addition to the gauge interaction (5) we have the Yukawa
couplings:
\beq
-{\it L}_Y=h_u
\bar{q}_{Li}\s_{ij}u_R\r^*_j+h_d\bar{q}_{Li}d_R\r^i+H.c.+(u,d,h_u,h_d\to
u',d',h'_u,h'_d).
\eeq

Here the Higgs field $\r^i=(\r^+,\r^0)$ and the masses of the fermions are:
 $m_u=h_uv/\sqrt{2}, \; m_d=h_dv/\sqrt{2},\; m'_u=h'_uv/\sqrt{2},\;
m'_d=h'_dv/\sqrt{2}$.

Using the fields $\psi$ and $\eta$ of eq.(4) one can rewrite the Lagrangian
(6) in the form:
\beq
-{\it L}_Y =\p_L M\eta_R +\e_R M^+\psi_L-\p_R\s M'^*\s\eta_L-\e_L\s M'^T
\s\psi_R,
\eeq
where the mass matrix $M(x)$ is
\beq
M(x)\; =\; \left( \begin{array}{cc}  h_u\r^0(x)^*,  &   h_d\r^+(x)\\
-h_u\r^+(x)^*,  &  h_d\r^0(x)  \end{array} \right),
\eeq
and $M'$ differs from $M$ by changing  $h_u,h_d\to h'_uh'_d$.

The interactions (5) and (7) can be rewritten together in the compact form,
using the eight-component spinor $\Psi$:
\beq
\Psi =\; \left( \begin{array}{c} \psi\\ \eta\end{array}\right).
\eeq
Then
\beq
L = \bar{\Psi} \hat{T} \Psi.
\eeq
Here
\beq
\hat{T} = \left( \begin{array}{cc} i\dd &-MR + \epsilon M'^* \epsilon L \\
-M^+ L + \epsilon M'^T \epsilon R & i\nn \end{array} \right) ,
\eeq
where $L,R =(1\pm \gamma_5)/2$.

Now we are ready to introduce the generalization of the usual $\gamma_5$
chirality:
\beq
\Gamma_5 = \gamma_5 (-1)^{2I+1},
\eeq
where $I$ is the weak isospin.

Obviously $\Gamma_5 = +1$
for the left-handed isospinor fermions (2) as well as for the right-handed
isoscalar fermions $u_R$, $d_R$, $u'_R$, $d'_R$.
On the other hand for the eight component fermion \f \ $\Psi$
(9), entering eq.(10), $\Gamma_5$ is
\beq
\Gamma_5 = \left( \begin{array}{cc} \gamma_5 & 0\\ 0& -\gamma_5
\end{array} \right).
\eeq

The matrix $\Gamma_5$ anti-commutes with the \op \ $\hat{T}$
of eq.(11) and with the matrix
\beq
\Gamma_0
= \left( \begin{array}{cc} \gamma_0& 0\\ 0& \gamma_0
\end{array} \right).
\eeq
Therefore
the lagrangian (10) is invariant under the generalized chiral
transformations:
\beq
\Psi \to e^{i\theta \Gamma_5 /2} \Psi,\qquad
\bar{\Psi} \to e^{i\theta \Gamma_5 /2} \bar{\Psi}.
\eeq

We have already mentioned that $\Gamma_5 = +1$ for both left-handed
and right-handed quarks $q$, $q'$. In other words for \f s
$q$, $q'$ the transformations (15) reduce to the phase rotations:
\beq
q \to e^{i\theta /2} q, \qquad q' \to e^{i\theta /2} q'.
\eeq

The symmetry under the transformations (15) is a classical symmetry which
is violated at the quantum level.
To see this we can intrioduce, as usually,
the regulator (eight-component) \f \ $\Psi_{reg}$ and to calculate
the famous
triangle diagram with two external $W$ lines and an insertion
of $i\theta M_{reg} \Gamma_5$ in the third vertex.
The result is usual:
the effective action acquires a contribution $\Delta S$,
defined by eq.(1), under the transformation (15).
This corresponds to nonconservation of the current $J_{\mu}^5$:
\beq
\partial_{\mu} J_{\mu}^5 = \frac{g_W^2}{16\pi^2}
W_{\mu\nu}^a \widetilde{W}_{\mu}^a ,
\eeq
$$J_{\mu}^5\; =\; \bar{\Psi} \gamma_{\mu} \Gamma_5 \Psi\; =\;
\bar{\psi} \gamma_{\mu} \gamma_5 \psi -
\bar{\eta} \gamma_{\mu} \gamma_5 \eta .$$
Coming back from the \f s $\psi ,\; \eta$
to $q, \; q'$ by eqs.(4) we see that the axial current $J_{\mu}^5$
is nothing but the vector baryon current (in accordence with (16)):
\beq
J_{\mu}^5 \;=\; B_{\mu}\; =\; \bar{q} \gamma_{\mu} q + \bar{q'} \gamma_{\mu}
q', \eeq
so that eq.(17) expresses baryon number violation.  Thus we arrive
to the conclusion that the weak \tt \ appears when the baryon number
rotation is performed, $\exp(iB\h/2)$, (i.e. the transformation of
eq.(16)) and that the appearence of the \tt \ is due to baryon number
nonconservation.

It is now easy to see that the effective action does not actually
depend on the $\theta$-parameter.
Suppose that the \tt \ is either introduced to the action from the
beginning or has appeared due to a certain chiral-type rotations
used in diagonalization of the mass matrix.
The effective action $\Gamma$ is expressed through the partition
function $Z$ by the following equation:
\beq
\Gamma\; =\; -  \ln Z_{tot} =\; - \ln \sum^{+\infty}_{Q_T = -\infty}
Z(Q_T).
\eeq
Here each of the terms $Z(Q_T)$ stands for the sector of the theory with
a certain integer topological charge $Q_T$:
\beq
Q_T\;=\; \frac{g_W^2}{32\pi^2}
\int d^4 x W^a_{\mu\nu} \widetilde{W}^a_{\mu\nu}.
\eeq
We use the euclidean formulation of the theory which is obtained by the
standard
procedure \cite{vzns}.
It is essential that under the Euclidean rotation $\bar{\Psi} \to
-i \Psi^+$, and the \f \ $\Psi^+$ is independent variable of integration
rather than the complex conjugated \f \ to $\Psi$.
The notation $\Psi^+$ means no more than that it transforms under
euclidean $O(4)$ rotations by a complex conjugated spinor representation.
However under the generalized chiral \rt \ (15)
$\Psi$ and $\Psi^+$ transform
as follows
\beq
\Psi\;\to\; e^{i\theta \Gamma_5 /2} \Psi, \qquad
\Psi^+ \to e^{i\theta \Gamma_5 /2} \Psi^+ .
\eeq
It is just this transformation which leaves invariant the classical
euclidean action.
The latter can be obtained from eqs.(10) and (11) by euclidean \rt :
\beq
S = \int d^4 x \Psi^+ \hat{T} \Psi,
\eeq
\beq
\hat{T} = \left( \begin{array}{cc} -i\dd & -iMR +
i\epsilon M'^* \epsilon L \\ -iM^+ L + i\epsilon M'^T \epsilon R& -i\nn
\end{array} \right).
\eeq
The quantity $Z(Q_T)$ is determined by the functional integration
over the Higgs and gauge \f s
\beq
Z(Q_T) = \int DW_{\mu}^a\; D\phi\; D\phi^* \times
\eeq
$$\times\; \exp \left[-\int d^4 x [1/4 W_{\mu\nu}^a\;W_{\mu\nu}^a +
|\nabla_{\mu} \phi|^2 + \lambda/2 (|\phi|^2 - v^2)^2 +\ln \Delta (W_{\mu}^a,
\phi)\right],$$
where the fermion determinant $\Delta$ reads as
$$\Delta  =  \int D\Psi \;D\Psi^+ \;\exp(-S).$$
We can now show that the effective action $\Gamma$, eq.(19),
is independent on the $\theta$-parameter.  To see this we change the
variables $\Psi$ and $\Psi^+$ in eq.(24) by the \rt \ (21) with a certain
$\theta =\alpha$.  Since we already know that this \trn \ is anomalous the
action will acquire an additional \tt \ with $\theta =\alpha$, i.e. the
initial $\theta$ changes: $\theta\to \theta +\alpha$.  That means that for
any $Q_T$
\beq Z(Q_T) = e^{i\alpha Q_T} Z(Q_T).
\eeq
So, we have
\beq
Z(Q_T) = 0 \;\;\;{\rm for} \;\; Q_T\neq 0.
\eeq
Thus for any nonvanishing topological charge the contribution $Z(Q_T)$
to the partition function
is zero.
Only the term with $Q_T =0$ remains in the sum in eq.(19).
For $Q_T =0$, however, the \tt \ (1) vanishes by definition.
We see therefore that the effective action $\Gamma$ does not depend on
$\theta$.

Certainly eq.(26) implies that the fermion determinant $\Delta$ vanishes
for $Q_T \neq 0$.
This means that quarks have zero modes in the external gauge \f \ with
$Q_T \neq 0$.
The existence of fermion \zm \ for the topologically nontrivial
configuration of the gauge and Higgs \f s has been discovered in
ref.\cite{rub}.  The explicit expression has been given for the fermion wave
function for the particular case of equal masses of up- and down- quarks.
Some more details concerning the \zm \ have been presented in our
previous paper \cite{aj}.
We shall rederive now the index theorem which has been formulated in \cite{aj}.

Consider the functional integral which differs from eq.(24) by insertion of
the several
factors of the type:
\beq
\Psi^+ \frac{1 \pm \Gamma_5}{2} \Psi.
\eeq
Obviously under the transformation (21) these factors transform as
\beq
\Psi^{+} \frac{1 \pm \Gamma_5}{2} \Psi\;\to\; e^{\pm i\theta}
\Psi^{+} \frac{1 \pm \Gamma_5}{2} \Psi.
\eeq
Let us insert $n_{+}$ factors $\Psi^+ \frac{1+\Gamma_5}{2} \Psi$
and $n_-$ factors $\Psi^+ \frac{1-\Gamma_5}{2} \Psi$
into the integral (24) for $Z(Q_T)$ and denote the resulting
integral by $Z(Q_T, n_+,n_-)$ ($Z(Q_T,0,0) = Z(Q_T)$).
Then under the change of variables given by eq.(21) with
$\theta =\alpha$ the Green function $Z(Q_T,n_+,n_-)$
acquires the phase factor
\beq
Z(Q_T, n_+,n_-)\; =\; e^{i\alpha (n_+ -n_- +Q_T)}
Z(Q_T, n_+,n_-).
\eeq
Therefore
\beq
Z(Q_T, n_+,n_-) \; \neq\; 0
\eeq
only if
\beq
n_+ - n_- \;=\; -Q_T.
\eeq
The integers $n_+$ and $n_-$ coincide with the numbers of the left-handed
($\Gamma_5 = +1$) and right-handed ($\Gamma_5 = -1$) \zm s of the
\op \ $\hat{T}$ defined by eq.(23).
Indeed consider for example the insertion of
the factor $\Psi^+ \frac{1+\Gamma_5}{2} \Psi$
in the functional integral over fermionic fields (see eq.(24))
\beq
\int  D\Psi \;D\Psi^+ \; \Psi^+ \frac{1+\Gamma_5}{2} \Psi \;\; e^{-S}.
\eeq
Expand $\Psi(x)$ in the eigen functions of the \op \ $\hat{T}$ defined by
eq.(23) \footnote{Strictly speaking the \op \ $\hat{T}$ is not Hermitian.
More accurate consideration is presented in ref.\cite{aj}.}
\beq
\Psi (x)\; =\; \sum_n c_n \Psi_n (x).
\eeq
Let $\Psi_0$ be a left-handed (for example) \zm \ of $\hat{T}$.
Then the action $S$ does not depend on $c_0$ and $c_0^+$ from
the expansion (33). The integral (32) is proportional to
\beq
\int c_0^+ dc_0^+\;\; c_0 dc_0 \;\neq\; 0.
\eeq
Thus the insertion of the factor $\Psi^+ \frac{1+\Gamma_5}{2} \Psi$
saves the integral from vanishing in integration over the Grassman variables
$c_0,\; c_0^+$.
If the \op \ $\hat{T}$ has $n_L$ left-handed and $n_R$ right-handed \zm s
it is necessary to insert $n_L$ factors $\Psi^+ \frac{1+\Gamma_5}{2} \Psi$
and $n_R$ factors $\Psi^+ \frac{1-\Gamma_5}{2} \Psi$
into the integral to have nonvanishing result.
We see that actually $n_+ =n_L$ and $n_- =n_R$, so that
\beq
n_L -n_R\; =\; -Q_T.
\eeq
This is the index theorem
which has been more rigorously proved in ref.\cite{aj}.

Thus contrary to the effective action $\Gamma$ the Green functions of
the type of $Z(Q_T, n_L,n_R)$ may depend on $\theta$-parameter.
As it was shown in \cite{aj} this dependence, however,
is unobservable since the amplitudes $Z(Q_T, n_L,n_R)\propto {\rm exp}
Q_T\theta$ do not interfere for different $Q_T = n_R -n_L$.

\section{Weak \tt \ Through Explicit Violation of Baryon Number}

We have seen that $\theta$ dependence
of the effective action is absent due to $\Gamma_5$ invariance of
the classical action. To have an ovbservable $\theta$ term the $\Gamma_5$
invariance should be violated explicitly.
On the other hand chiral \rt \ (15) is nothing but the phase \trn \
(16) generated by the baryon charge.
Therefore to violate the $\Gamma_5$ invariance
it is necessary to violate the baryon number conservation.
For one generation of fermions the simplest effective
Lagrangian violating $B$ but conserving electric and colour has the form:
\beq
{\it L}\; =\; G \epsilon^{ijk} (q_{iL}C\epsilon q_{jL}) (q_{kL}C\epsilon l_L)
+ {\rm H.c.},
\eeq
where $i,j,k$ are the colour indices, $G$ is the effective coupling constant.
We use only the \f s with left-handed $\gamma_5$ chirality since only
left-handed fermions interact with $W$ bosons.
Of course in the presence of fermion masses (Yukawa couplings)
right-handed \f s can also be used.

For three generations of quarks and leptons the effective interaction should
be a product of the \op s of the type of eq.(36).
This product is the same as the t'Hooft effective Lagrangian induced by
instantons \cite{1} but in this case the interaction
is introduced explicitly.

Certainly the Lagrangian (36) is nonrenormalizable but one can understand
it as a low-energy limit for a renormalizable interaction.
For instance:
\beq
{\it L} \;=\; h_1 \epsilon^{ijk} (q_{iL}C\epsilon q_{jL}) \Phi_k +
h_2 \Phi^{i*}\epsilon (q_{iL}C\epsilon l_L) +{\rm H.c.},
\eeq
where $\Phi_k$ is a scalar colour-triplet heavy \f .
Integarting out $\Phi_k$ we get the Lagrangian (36).

Obviously in the real world the coupling constant $G$ should be very small
and this leads to the additional suppression of the $\theta$ dependence
of the effective action.
It is easy to understand the structure of this dependence.
Perturbatibely in expansion in $G$ the Lagrangian (36) and its Hermitian
conjugated are inserted into the functional integral and therefore leads
to non-vanishing contribution when integrated over $c_0$ and $c_0^+$ (the
Grassman variables corresponding to fermion \zm s).  One obtains:
\beq
\Gamma=-\ln \sum Z(Q_T) = const+N\left[(Gv^2)^2
e^{-\frac{8\pi^2}{g_W^2} -
i\theta}+(Gv^2)^2 e^{-\frac{8\pi^2}{g_W^2} + i\theta}\right],
\eeq where $N$
is the dimensionless factor of order of unity ($N$ however may contain the
powers of the coupling constant $g_W$), $v$ is the v.e.v. of the standard
Higgs \f .  The term proportional to exp$(-i\theta)$ comes from the sector
$Q_T =-1$ while the complex conjugated term corresponds to $Q_T =+1$.

\section{Weak \tt \ in Left-Right Symmetric Generalization of
The Standard Model}

We shall determine now the $\h$ dependence of the effective action
(vacuum energy) in the \lr \ model.
This model contains $W^L_{\mu}$ and $W^R_{\mu}$ gauge bosons
corresponding to the $SU(2)_L$ and $SU(2)_R$ groups.
The simplest Yukawa couplings of the model are constructed by using
the Higgs \f \ $\chi^{\beta}_{\alpha} = (1/2,1/2^*)$.
For the quark doublets $(u_L,d_L) = (1/2,0)$ and $(u_R,d_R) = (0,1/2)$
one has
\beq
-{\it L}\;=\; h_1 \bar{q}^{\alpha}_L q_{R \beta}\chi^{\beta}_{\alpha} +
h_2 \bar{q}^{\alpha}_L q_{R \beta}(\epsilon
\chi^*\epsilon)^{\beta}_{\alpha}+{\rm H.c.} =
\bar{q}^{\alpha}_L M^{\beta}_{\alpha}q_{R \beta}+{\rm H.c.},
\eeq
where
\beq
M = \left( \begin{array}{cc} h_1\chi^1_1 -
h_2\chi^{2*}_2& h_1\chi^2_1 +h_2\chi^{1*}_2 \\
h_1\chi^1_2 +h_2\chi^{2*}_1 & h_1\chi^2_2 -
h_2\chi^{1*}_1 \end{array} \right).
\eeq
The vacuum expectation values of the neutral Higgs bosons
$<\chi^1_1> = v_1\;$ and $<\chi^2_2> = v_2$
($v_1,\;v_2$ are real, for simplicity) give masses to the $u$- and
$d$- quarks:
\beq
m_u = h_1v_1 -h_2v_2, \qquad  m_d = h_1v_2 -h_2v_1.
\eeq
The model can include also several other Higgses which do not interact with
fermions but contribute to the masses of $W_L$ and $W_R$ bosons.

Constructing eight-component spinors, as it has been done in sect.2,
the fermion Lagrangian can be written in the form of eqs.(22)-(23)
but with the modified \op \ $\hat{T}$:
\beq
L = \Psi^+ \hat{T} \Psi, \qquad  \Psi=(\psi,\eta),
\eeq
\beq
\hat{T} = \left( \begin{array}{cc} -i\dd_L &-iMR + i\epsilon M'^* \epsilon L\\
-iM^+L + i\epsilon M'^T \epsilon R& -i\dd_R \end{array} \right).
\eeq
Here $\dd_{L,R} = \gamma_{\mu} D_{\mu L,R}$ where $D_{\mu L}$
and $D_{\mu R}$ are the covariant derivatives including $W_{\mu L}$ and
$W_{\mu R}$
bosons, respectively.
$M$ and $M'$ are constructed according to eq.(40).

The \op \ $\hat{T}$ remains anti-commuting with $\Gamma_5$ as defined by
eq.(13).
Therefore the classical Lagrangian is invariant under the
generalized chiral transformations (15).
The anomalous divergence of the baryon current $J_{\mu}^5$
(eqs.(17) and (18)) is now
\beq
\partial_{\mu} J_{\mu}^5 \; = \;
\frac{g_L^2}{16\pi^2} W^{La}_{\mu\nu} \widetilde{W}^{La}_{\mu\nu}
- \frac{g_R^2}{16\pi^2} W^{Ra}_{\mu\nu} \widetilde{W}^{Ra}_{\mu\nu}.
\eeq
Changing the variables $\Psi$ and $\Psi^+$
by eq.(21) with $\theta =\alpha$ we get instead of eq.(25)
\beq
Z(Q_T^L ,Q_T^R)\; = \; e^{i\alpha (Q_T^L -Q_T^R)}
Z(Q_T^L ,Q_T^R),
\eeq
where $Q_T^L$ and $Q_T^R$ are the corresponding topological charges of the
$W_{\mu L}$ and $W_{\mu R}$ \f s.

Thus it is possible that $Z(Q_T^L ,Q_T^R)\neq 0$
only if $Q_T^L =Q_T^R$. The generalization of the index theorem (35) reads:
\beq
n_L -n_R\; =\; -(Q_T^L - Q_T^R).
\eeq
Here $n_L$ and $n_R$ are the numbers of the left-handed $(\Gamma_5 =+1)$
and right-handed $(\Gamma_5 =-1)$
fermion \zm s of the \op \ (43).

Consider the case $Q_T^L = Q_T^R =-1$.
In the absence of the Yukawa couplings there is one left-handed \zm \
$(\Gamma_5 =+1,\;\;\gamma_5 =+1)$, $n_L = +1$,
and one "right-handed" \zm \, $n_R =+1$
($\Gamma_5 =-1$ while $\gamma_5 =+1$).

In the presence of the Yukawa couplings one can expect, however according
to eq.(46),
that
$n_L = n_R =0$
and there are no \zm s at all.
We shall see that this is indeed the case by calculation of the
partition function (vacuum energy)
using the explicit 't Hooft's solution \cite{1}.
The partition function does not vanish in the sector $Q_T^L =Q_T^R =-1$
which means the absence of \zm s.

The vacuum energy $\epsilon V_4$ ($\epsilon$ is the energy density, $V_4$
is the Euclidean 4-volume) is given by
\beq
\epsilon V_4\; =\; - \ln \sum_{Q_T^L ,Q_T^R}
Z(Q_T^L ,Q_T^R),
\eeq
where the contributions to the partition function $Z(Q_T^L ,Q_T^R)$,
corresponding to different values of the topological charges,
are determined by the equations similar to eqs.(24).
The difference with eq.(24) is that there are now integrations over
$W^L_{\mu}$ and $W^R_{\mu}$
\f s as well as over all the Higgs \f s included in the model.

The $\theta$ dependence of $\epsilon V_4$ comes from the sector with
$Q_T^L =Q_T^R =Q_T \neq 0$.
We leave
only the contributions $Q_T = \pm 1$ since they have minimal exponential
suppression. One has
\beq
\epsilon V_4 \;= \; -\frac{Z(+1) +Z(-1)}{Z(0)} + const.
\eeq
We are now going to determine the dependence of the vacuum energy on $\theta$
parameter.

In the rest of this paper we consider the realistic case with 12 weak fermion
doublets (3 generations of quarks and leptons).
As it has been done in sect.2
we combine the pairs of the Weyl doublets into the Dirac doublets
$\psi_k ,\;\;\eta_k$, $k= 1,...,6$, where $\psi$ and $\eta$ transform
as $\psi=(1/2,0)$ and $\eta = (0,1/2)$.
The Yukawa couplings are represented now by the mass matrices of the type
of eq.(40) but with the different coupling constants for different fermions.
So there are six $M_k(x)$ and six $M'_k(x)$ matrices.
The Yukawa couplings are necessary to compensate each of the 6 \zm s
of $\psi_k$ and 6 \zm s of $\eta_k$ (and their complex conjugated) which
exist in the case of purely massless fermions.
To the leading nonvanishing approximation in the Yukawa
interactions the vacuum energy has the form (see the form of
the Yukawa couplings from eq.(43)):
\begin{eqnarray}
\epsilon V_4& =& - e^{-2i\theta} <\prod^6_{k=1} \int  d^4 x_k d^4 y_k
(\psi^+_{Lk}(x_k) \epsilon M'^*_k (x_k)
\epsilon \eta_{Lk} (x_k)) \nonumber \\
&\times & (\eta^+_{Lk} (y_k)M^+_k (y_k)\psi_{Lk}(y_k))>\; +\;{\rm  H.c.}
\end{eqnarray}
Here $<...>$ implies the functional integration over all the \f s
of the model with the weight $exp(-S)$
where $S$ is the action,
without Yukawa couplings.
The factor $\exp(-2i\theta)$ corresponds to the \tt \
\beq
\theta \left[ \frac{g_L^2}{32\pi^2} \int d^4 x W^{La}_{\mu\nu}
\widetilde{W}^{La}_{\mu\nu} + \frac{g_R^2}{32\pi^2} \int d^4 x W^{Ra}_{\mu\nu}
\widetilde{W}^{Ra}_{\mu\nu}\right]
\eeq
for $Q_T^L =Q_T^R =-1$.
The complex conjugated term in eq.(49) corresponding to $Q_T^L =Q_T^R =+1$
contains ${\rm exp}(+2i\theta)$ factor
and right-handed fermion modes $\psi_{Rk},\;\; \eta_{Rk}$.

To calculate the integrals (49) we use the anti-instanton gauge \f s
in the singular gauge (the notations are the same as in ref.\cite{vzns}):
\beq
W^L_{\mu,antiinst} =
\frac{2\rho^2_L}{g_L} \frac{\eta_{a\mu\nu}
\tau_a (x-x_L)_{\nu}}{(x-x_L)^2 [(x-x_L)^2 + \rho^2_L]},
\eeq
$$W^R_{\mu,antiinst} =
\frac{2\rho^2_R}{g_R} \frac{\eta_{a\mu\nu}
\tau_a (x-x_R)_{\nu}}{(x-x_R)^2 [(x-x_R)^2 + \rho^2_R]}.$$
Here $x_L,\;x_R$ and $\rho_L,\;\rho_R$ are the coordinates of the centres and
the sizes of the instantons.
One should integrate also over the orientations of the instantons, i.e.
actually we use instead of eqs.(51):
\beq
W^{L}_{\mu,antiinst} \to U_L W^L_{\mu, antiinst}U^+_L ,\;\;\;
W^{R}_{\mu,antiinst} \to U_R W^R_{\mu, antiinst}U^+_R
\eeq
and integrate over the constant matrices $U_L$ and $U_R$.

After the integration over the Grassman variables
corresponding to the fermion \zm s,
all the \op s $\psi_{Lk} (x_k)$ and $\eta_{Lk} (x_k)$ are
changed to the \zm \ wave functions (c-numbers).
These wave functions have the form:
\begin{eqnarray}
\psi^{il}_{0L}&=& \frac{1}{\pi\sqrt{2}}
\frac{\rho_L}{[(x-x_L)^2 + \rho_L^2]^{3/2}}
\left( \frac{i(x-x_L)_{\mu}\tau_{\mu}^+
\epsilon }{\sqrt{(x-x_L)^2}}\right)^{il} , \\
\eta^{il}_{0L}&=& \frac{1}{\pi\sqrt{2}}
\frac{\rho_R}{[(x-x_R)^2 + \rho_R^2]^{3/2}}
\left( \frac{i(x-x_R)_{\mu}\tau_{\mu}^+
\epsilon }{\sqrt{(x-x_R)^2}}\right)^{il},\nonumber
\end{eqnarray}
$$\epsilon = i\tau_2.$$
Here $i,\;l$ indices refer to the isospin and the usual spin, correspondingly.
The expressions for the wave functions (53) are fixed for the "unrotated"
anti-instanton \f s (51).
Under the \rt \ (52):
\beq
\psi_{0L} \to U_L \psi_{0L},\qquad \eta_{0L} \to U_R \eta_{0L}.
\eeq
A slightly misleading notation in the second eqs.(53) and (54) are the
result that the left-handed (in the usual sense, $\gamma_5 =+1$)
\zm \ $\eta_{0L}$ is right-handed in a sense that $\Gamma_5=-1$.
It corresponds to the anti-instanton gauge \f \ $W^R_{\mu, antiinst}$
of the $SU(2)_R$ group.

To compute the vacuum energy (49)
we should also find the extremum configuration for the Higgs \f s and
the mass matrices $M_k(x_k)$ and $M'_k(x_k)$.
The exact problem which we face now is to solve the equation for the Higgs \f \
$\chi$
\beq
D^2 \chi\;=\;0,
\eeq
where the covariant derivative $D_{\mu}$ reads now
\beq
D_\mu\; =\; \partial_\mu - ig_L W^{La}_{\mu} t^a_L
- ig_R W^{Ra}_{\mu} t^a_R.
\eeq
Here $t^a_L$ and $t^a_R$ are the generators of the $SU(2)_L$ and
$SU(2)_R$
groups and the gauge \f s are to be taken from eqs.(51) or (52).

For our purposes we shall consider
the situation corresponding to the dilute
instanton-antiinstanton gas when
\beq
|x_L - x_R|\; \gg\; \rho_L,\; \rho_R.
\eeq
Though this is not justified by the smallness of any physical parameter
it seems that the factorized form of the approximate solution which
is given below can not lead to a big error for the integral (49).
Using 't Hooft's solution \cite{1,vzns} one can easily find
under the conditions (57)
\beq
\chi \;=\; f_1 f_2 \left( \begin{array}{cc}
v_1&0\\ 0& v_2 \end{array} \right) \;=\;
 f_1 f_2 \chi_0,
\eeq
where the factors $f_1f_2$ are
\beq
f_1 = \sqrt{\frac{(x-x_L)^2}{(x-x_L)^2 + \rho_L^2}}, \qquad
f_2 = \sqrt{\frac{(x-x_R)^2}{(x-x_R)^2 + \rho_R^2}}.
\eeq
In the vicinity of the center of each instanton only one of the factors
$f_1$ or $f_2$
survives while the other is approximated by unity and we pass to 't Hooft's
solution.

One get from eqs.(58) and (59) for the $M(x)$ matrix for the case of, say,
$(u,d)$ quarks
\beq
M(x) = f_1 f_2 M_0, \qquad  M_0=\left( \begin{array}{cc} m_u & 0 \\ 0 & m_d
\end{array} \right).  \eeq
The important point about this solution is that in contrast to the
case of the fermion \zm s it is valid not only for the "unrotated"
instanton \f s (51) but also for fields given in eq.(52).
This is due to the fact that though under the \rt \ of
the gauge \f s (52) the matrix $\chi(x)$ in eq.(55) changes to
$\chi \to U_L \chi U_R^+$
the new matrix $ \chi'(x) = f_1(x) f_2(x)
U_L \chi_0 U_R^+$ still satisfies eq.(55) since the constant matrix
$U_L \chi_0 U_R^+$
drops out of this equation.
Therefore the solution (60), which is fixed by the requirement that
$M(x) \to M_0$ at $|x| \to \infty$ and can not be rotated itself,
refers nevertheless to the arbitrary orientaion
of the instanton \f s.
That means that the orientaion
of the fermion \zm s in the isotopic space in the integrals (49)
(which follows the orientaion of the instanton
\f s) is not correlated to the orientaion of $M^i_j(x)$ and $M'^i_j(x)$,
which are to be taken in the form (60).

We are ready now to write down each the factors in eq.(49).
Namely:
\begin{eqnarray}
I'_k &=& \int \frac{d^4 x_k}{2\pi^2}
\frac{\rho^{3/2}_L}{[(x_k -x_L)^2 + \rho^2_L]^2}
\frac{\rho^{3/2}_R}{[(x_k -x_R)^2 + \rho^2_R]^2} \times \nonumber\\
&\times& Tr[(x_k-x_L)_{\mu} \tau_{\mu}^- U^+_L \epsilon
M'^*_{0k} \epsilon U_R (x_k-x_R)_{\nu} \tau_{\nu}^+],
\end{eqnarray}
\begin{eqnarray*}
I_k &=& \int \frac{d^4 y_k}{2\pi^2}
\frac{\rho^{3/2}_L}{[(y_k -x_L)^2 + \rho^2_L]^2}
\frac{\rho^{3/2}_R}{[(y_k -x_R)^2 + \rho^2_R]^2} \times \\
& \times & Tr[(y_k-x_R)_{\mu} \tau_{\mu}^- U^+_R
M^+_{0k} U_L (y_k-x_L)_{\nu} \tau_{\nu}^+].
\end{eqnarray*}
The additional factors
$\sqrt{\rho_L}$ and $\sqrt{\rho_R}$
in the fermion \zm s in these equations,
as compared to eqs.(53), come out of
the correct normalization at the integartion
over the \zm s \cite{vzns}.

The total expression for the vacuum energy can be obtained
now by using the instanton measure
\begin{eqnarray}
\epsilon V_4& =& -c^2 e^{-2i\theta} \int d^4 x_L d^4 x_R
\frac{d\rho_L}{\rho^5_L}\frac{d\rho_R}{\rho^5_R}
e^{-\frac{8\pi^2}{g^2_L(\rho_L)}- \pi^2 v_L^2 \rho^2_L} \\
&\times& e^{-\frac{8\pi^2}{g^2_R(\rho_R)}- \pi^2 v_R^2 \rho^2_R}
\left(\frac{8\pi^2}{g^2_L(\rho_L)}\right)^4
\left(\frac{8\pi^2}{g^2_R(\rho_R)}\right)^4
\int dU_L dU_R \prod^6_{k=1} I'_k I_k +{\rm H.c.}\nonumber
\end{eqnarray}
Here $v_L$ and $v_R$ may differ from $v_1 = <\chi^1_1>$ and
$v_2 = <\chi^2_2>$
since $v_L$ and $v_R$ are connected to the masses of $W_L$ and $W_R$,
i.e. to all the Higgses of the theory, not only those interacting
with fermions; $g^2_L(\rho_L)$ and
$g^2_R(\rho_R)$ are the running gauge coupling constants for the $SU(2)_L$ and
$SU(2)_R$ groups at
the scales $\rho_L$ and $\rho_R)$ respectively.
The numerical constant $c$ is
\beq
\left.c\;= \; 0.6419\; \times \; 1.157^{N_F-N_H}\right/4\pi^2,
\eeq
where $N_F$ is the number of the fermion doublets of the $SU(2)_L$
(or $SU(2)_R$)
group ($N_F = 12$), $N_H$ is the number of the Higgs
doublets
(say, $N_H =2$ if only two doublets of $SU(2)_L$ group
entering $\chi^{\beta}_{\alpha}$ are taken into account).

The $4\pi^2$ factor in eq.(63) comes from the \zm s,
together with the well-known factors $(8\pi^2 /g^2_{L,R})^4$,
while the other factors are related to
non-zero modes.
The integrals over the orientations $U_L$ and $U_R$
are normalized to unity.

The further calculation of the integrals in eq.(62) are
given in Appendix.
We present here the final result for the energy density
\beq
\epsilon\;=\; - 0.122 \frac{g_R^3}{g_L^{11}}
\frac{M^3(W_L)}{M^{11}(W_R)}
m_e m_{\mu} m_{\tau} m^3_d m^3_c m^3_t\;\cos 2\theta \times
\eeq
$$\times {\rm exp}\;(-\frac{8\pi^2}{g^2_L}-\frac{8\pi^2}{g^2_R}).$$
Here the gauge coupling constants $g_L$ and $g_R$ are taken at the
scales $v_L$ and $v_R$ respectively.
Numerically the value of $\epsilon$ is very small.
Taking
$g_L =g_R = 0.637$ one has for the exponential suppression:
\beq
{\rm exp}\;(-\frac{8\pi^2}{g^2_L}-\frac{8\pi^2}{g^2_R})\;
= \;10^{-169}
\eeq
while the other factors give for, say,
$M(W_R)= 300 \;\gv$ and $m_t =140 \;\gv$,
the value $10^{-24} \;\gv^{-4}$.
So one has
\beq
\epsilon = - 10^{-193}
\cos\; 2\theta\; \gv^4.
\eeq
Obviously the smallness of that type is typical for the $\theta$
dependence of any amplitudes with baryon nonconservation
where it could manifest itself by $CP$ violation.
However since it is not excluded that the exponential
suppression might disappear at multi-TeV energies
the problem of \tt \ for the \lr \ theories could be not completely
an academical one.

Another question which has been raised before in connection to
some astrophysical speculations \cite{aa},
is the question
of the mass of the arion --- "massless"
axion, a Goldstone boson not coupled to the strong anomaly
\cite{au}.
The question is what value of the mass can have arion related to
the weak anomaly.
Equation (66) immediately gives the
answer to this question.
Actually arion is nothing but $\theta$ angle when it is made to be a
dynamical degree of freedom ($\theta = constant \to \theta =\theta (x)$)
by widening of the Higgs sector. The normalized arion \f \ $a(x)$
differs from $\theta (x)$
by a constant factor, $a(x) = V \theta (x)$,
where $V$ is the scale of the symmetry breaking accosiated with arion.
Typical value of $V$ is $V = 10^{10} \gv$.
That leads to the mass of the arion
\beq
m_a \; = \; \frac{10^{-193/2}}{10^{10}}\, \gv\; =\; 10^{-98}{\rm  eV}.
\eeq
This tiny value is much smaller than what has been discussed in ref.\cite{aa}
($m_a \sim 10^{-31} eV$ corresponding to $\hbar /m_a c \sim 100 Mpc$).

\section{Acknowlegments}

We are very grateful to C.Callan whose question at the seminar at Priceton
triggered this investigation.
One of us (A.A.) would like to thank Physics Department of
Princeton University and especially A.Polyakov for their hospitality.
A.J. is grateful to Theoretical Physics Institute of University of
Minnesota for the hospitality.

\appendix
\section{Appendix}
\renewcommand{\theequation}{A.\arabic{equation}}
\setcounter{equation}{0}

Consider first one of the factors $I'_k$ or $I_k$ in eqs.(61).
The products of the $\tau^{\pm}$ matrices entering the traces can be
simplified as follows
\beq
\tau_{\nu}^+ (x_k - x_R)_{\nu} (x_k - x_L)_{\mu}\tau_{\mu}^-
\to (x_k - x_R)_{\mu}(x_k - x_L)_{\mu},
\eeq
and analogously for the matrices in $I_k$.
The reason is that antisymmetric part of $\tau_{\mu}^-\tau_{\nu}^+$
($\tau_{\mu}^-\tau_{\nu}^+ =\delta_{\mu\nu} +i\eta_{a\mu\nu} \tau_a$)
can not survive after
the integration over $x_k$
since it depends only on a single vector $(x_L -x_R)_{\mu}$.

We obtain
\beq
I'_k = B \; Tr (V^+ \epsilon M'^*_{0k} \epsilon),\;\;\;
I'_k = B \; Tr (V M^+_{0k}),
\eeq
where
\beq
V= U_L U^+_R,
\eeq
$$B = \int \frac{d^4 x}{2\pi^2}
\frac{\rho^{3/2}_L \rho^{3/2}_R
(x-x_L) (x-x_R)}{[(x-x_L)^2 +\rho^2_L]^2 [(x-x_R)^2 +\rho^2_R]^2}.$$
The factor $B$ can be calculated by the usual Feynman prescriptions:
\beq
B= \frac{\rho^{3/2}_L \rho^{3/2}_R}{8}
\frac{\partial^2}{\partial x_{0\mu}\partial x_{0\mu}}
\int^1_0 dt \;{\rm ln} [x^2_0 t(1-t) +t \rho^2_L + (1-t) \rho^2_R]=
\eeq
$$= \;\frac{\rho^{3/2}_L \rho^{3/2}_R}{2}
\int^1_0 dt \;\frac{x_0^2 t^2 (1-t)^2 + 2 t^2 (1-t) \rho^2_L +2 t
(1-t)^2 \rho^2_R}{[x_0^2 t (1-t) + t \rho^2_L +
(1-t) \rho^2_R]^2},$$  $$ x_0 =x_L-x_R.$$

The last integral is very simple if $\rho_R \to0$ (or $\rho_L \to0$)
and equals to
$1/2\; \rho^{3/2}_L \rho^{3/2}_R  (x_0^2 + \rho^2_L)^{-1}$.
In the case where $\rho_L =\rho_R$ the expansion in $\rho^2_{L,R} /
x^2_0$ coincides with the expansion of the simple interpolation:
\beq
B\; =\; \frac{1}{2} \frac{\rho^{3/2}_L \rho^{3/2}_R}{x^2_0 +
\rho^2_L +\rho^2_R}.
\eeq
Since in any case we have used for the Higgs \f \ the approximate
solution which is valid only if $x^2_0 \gg \rho^2_L,\;\rho^2_R$
the expression (A.5) seems to be quite
satisfactory.

Next we need to calculate the integral
\beq
A \; =\; \int dV \prod^6_{k=1}
Tr(V^+ \epsilon M'^*_{0k}\epsilon ) Tr(VM^+_{0k}).
\eeq
(We remind
that $V = U_L U^+_R$ and so
$dU_L dU_R \to dV$).

By changing the variables one can see that the integral
(A.6) is invariant under the transformations
$M^+_{0k} \to U' M_{0k}^+$ and
$\epsilon M'^*_{0k}\epsilon \to \epsilon M'^*_{0k}\epsilon U'^+$
where $U'$ is an arbitrary unitary matrix.
Obviously,
the result of the integration in (A.6) is the trace of the
product of the factors
$\epsilon M'^*_{0k}\epsilon$
and $M^+_{0k}$.
The above mentioned invariance shows that the only
possible combinations are those where the factors $\epsilon M'^*_{0k}\epsilon$
and $M^+_{0k}$ follows one by another, i.e. of the type
$$Tr M^+_{01}(\epsilon M'^*_{01}\epsilon)
M^+_{02}(\epsilon M'^*_{02}\epsilon)...$$
Since the matrix $M_0$ is
$$\left( \begin{array}{cc} m_u & 0\\ 0 & m_d \end{array} \right),$$
and the matrix $\epsilon M_0\epsilon$
is
$$\epsilon M_0\epsilon\; =\; -\left( \begin{array}{cc}
 m_d& 0\\ 0&m_u \end{array}
\right),$$
it is easy to realize that the result for the integral (A.6)
should contain 6 masses of the upper components of the doublets and
6 masses of the lower components.
It is also evident that the result is symmetric over all doublets.
That fixes uniquely the structure of the integral (A.6):
\beq
A\; =\; N \sum_P m_{u_1}...m_{u_6}m_{d_1}...m_{d_6},
\eeq
where $N$ is a numerical factor.

In this equation we denote by $u_1...u_6 (d_1...d_6)$
certain 6 upper (down) components of the 12 doublets.
$\sum_P $ implies the sum over all possible permutations.
In the realistic situation one of the terms in (A.7)
is the largest.
To avoid neutrino masses one should use $m_e m_{\mu} m_{\tau}$
for 3 "down" lepton masses.
Then for the remaining quark doublets the mass of $d$-quark
is larger than the mass of $u$-quark.
That fixes the largest term in (A.7):
\beq
m_e m_{\mu} m_{\tau} m^3_d m^3_c m^3_t\; = \;\bar{m}_F^{12}.
\eeq
We denote here by $\bar{m}_F$ "the average" fermion mass.
The understanding
of the general structure of (A.7) allows to find
the numerical factor $N$.
Using (A.7) we can calculate $N$ putting all the masses equal unity.
One has from (A.6) and (A.7)
\beq
N C^k_{2k} =
\int dV (Tr V^+)^k (Tr V)^k.
\eeq
This equation is written down for arbitrary $k$,
actually $k=6$ and $C^k_{2k} =924$.
We have omitted in eq.(A.9) the factor $(-1)^k$ but for the general case
it is also present in eq.(49) and $(-1)^{2k} =+1$.

The integral (A.9) can be readily calculated if one parametrizes
$V = n_4 + in_i \tau_i,\;\; i=1,2,3$,
$n^2_4 + n^2_i =1$.
One gets
\beq
N \frac{(2k)!}{(k!)^2}\; = \; \frac{2^{2k+1}}{\pi}
\int d\alpha \; \sin^2 \alpha
\cos^{2k}\alpha = \frac{2(2k -1)!}{(k-1)! (k+1)!}.
\eeq
Or:
\beq
N\; = \; \frac{1}{k+1}.
\eeq
Collecting (A.2),(A.5),(A.7) and (A.11) we
obtain for the density of the vacuum energy from eq.(62)
\begin{eqnarray}
\epsilon & = &
-c^2 e^{-2i\theta} \int \frac{d\rho_L}{\rho_L^5}
\frac{d\rho_R}{\rho_R^5}
\left( \frac{8\pi^2}{g_L^2 (\rho_L)}\right)^4
\left( \frac{8\pi^2}{g_R^2 (\rho_R)}\right)^4 \times \nonumber \\
&\times & {\rm exp}(-\frac{8\pi^2}{g_L^2 (\rho_L)} -\pi^2 v_L^2 \rho^2_L
 -\frac{8\pi^2}{g_R^2 (\rho_R)} -\pi^2 v_R^2 \rho^2_R) \nonumber\\
&\times & \frac{1}{k+1} \; \frac{1}{2^{2k}}\;
\frac{\pi^2}{2(k-1)(2k-1)}
\frac{\rho_L^{3k} \rho_R^{3k}}{(\rho_L^2 +\rho_R^2)^{2(k-1)}}
\bar{m}_F^{2k} +{\rm H.c.}
\end{eqnarray}
We have integrated in eq.(A.12) over the positions of the
instanton centres.
Note that while one of the integrations gives the normalization volume
the other is divergent at $|x_L -x_R|\to \infty$ if $k=1$.
Therefore we must conclude that for the simplest case of two weak doublets
$(k=1)$ the calculation is not selfconsistent.
In this case the integrals should be cut off at large distances
of order of $M_W^{-1}$, where all the solutions decrease exponentially.
The value $M_W^{-1}$ has been assumed to be infinite ($M_W =0$)
at present calculations.

To do the remaining integrations over $\rho_L$ and $\rho_R$
we have to express the running coupling constants
$g^2_L(\rho_L)$ and $g^2_R(\rho_R)$
through $g^2_L(v_L)$ and $g^2_R(v_R)$
normalized at the scales $v_L$ and $v_R$.
Then, for instance:
\beq
e^{-8\pi^2/g^2_L(\rho_L)} \to (\rho_L v_L)^{b_1} e^{-8\pi^2/g^2_L(v_L)} ,
\eeq
where $b_1$ is the first coefficient of the Gell-Mann-Low
function (and the same, of course, for $g^2_R(\rho_R)$).
Numerically, $b_1 = 22/3 - n_f/3 - n_H/6$ where $n_F$ is the number of
the weak doublets and $n_H$ is the number of the Higgses.
For $n_F =12$ and $n_H =2$, $b_1=3$.
Then we have for the integral over $\rho_L$ (we put now $k=6$):
\beq
v_L^3 \int^\infty_0 d\rho_L
\frac{\rho_L^{16} e^{-\pi^2 v^2_L \rho^2_L}}{(\rho^2_L +
\rho^2_R)^{10}}.
\eeq
This integral converges at $\rho_L \sim \rho_R \sim v^{-1}_R \ll v^{-1}_L$
(since for the realistic case $M(W_L) \ll M(W_R)$).
Therefore we can neglect the exponential factor in
(A.14) and get for the integral
\beq
\frac{715\pi}{2^{17}}
\frac{v^3_L}{\rho^3_R}.
\eeq
The latter integral over $\rho_R$ is also easily calculated
\beq
v^3_L v^3_R \int^\infty_0
d\rho_R \;\rho_R^{16} e^{-\pi^2 v^2_R \rho^2_R} =
\frac{6!}{2\pi^{14}}
v^3_L v^{-11}_R.
\eeq
Gathering
all the factors together we obtain for $\epsilon$:
\begin{eqnarray}
\epsilon
&=& - \frac{585}{7}\, \frac{1}{(4\pi)^{15}}\, (0.6419)^2 (1.157)^{20}\,
\left(\frac{8\pi^2}{g_L^2 (v_L)}\right)^4
\left(\frac{8\pi^2}{g_R^2 (v_R)}\right)^4 \nonumber \\
& \times & e^{ -8\pi^2/g_L^2 (v_L)\,
 -\, 8\pi^2/g_R^2 (v_R)}
\frac{\bar{m}_F^{12} v_L^3}{v_R^{11}}\, \cos 2\theta.
\end{eqnarray}
Rewriting this eq. in terms of the masses
of $W_L$ and $W_R$ bosons we arrive at the expression
given by eq.(64).

\end{document}